\documentclass[preprint2]{aastex61}

\usepackage[english]{babel}
\usepackage[utf8x]{inputenc}
\usepackage[T1]{fontenc}
\usepackage{aas_macros}

\usepackage{graphicx} %allows you to use jpg or png images. PDF is still recommended
\usepackage{amssymb}  %

\begin{document}

\title{Rogue planets and brown dwarfs: Predicting the populations of free-floating planetary mass objects observable with JWST}

\email{as110@st-andrews.ac.uk}

\author[0000-0001-8993-5053]{Aleks Scholz}
\affiliation{SUPA, School of Physics \& Astronomy, University of St Andrews, North Haugh, St Andrews, KY16 9SS, United Kingdom}

\author[0000-0002-7989-2595]{Koraljka Muzic}
\affiliation{CENTRA, Faculdade de Ci\^{e}ncias, Universidade de Lisboa, Ed. C8, Campo Grande, 1749-016 Lisboa, Portugal}

\author[0000-0001-5349-6853]{Ray Jayawardhana}
\affiliation{Department of Astronomy, Cornell University, Ithaca, NY 14853, USA}

\author{Lyra Quinlan}
\author[0000-0003-0688-5332]{James Wurster}
\affiliation{SUPA, School of Physics \& Astronomy, University of St Andrews, North Haugh, St Andrews, KY16 9SS, United Kingdom}

\begin{abstract}
Free-floating (or rogue) planets are planets that are liberated (or ejected) from their host systems. Although simulations predict their existence in substantial numbers, direct observational evidence for free-floating planets with masses below $\sim 5\,M_{\mathrm{Jup}}$ is still lacking.  Several cycle-1 observing programs with JWST aim to hunt for them in four different star-forming clusters. These surveys are designed to be sensitive to masses of $1-15\,M_{\mathrm{Jup}}$ (assuming a hot-start formation), which corresponds to spectral types of early L to late T for the ages of these clusters. If the existing simulations are not wide off the mark, we show here that the planned  programs are likely to find up to 10-20 giant rogue planets in moderate density clusters like NGC1333 or IC348, and several dozen to $\sim 100$ in high-density regions like NGC2024 and the Orion Nebula Cluster. These numbers correspond to 1-5\% of the total cluster population; they could be substantially higher if stars form multiple giant planets at birth. In contrast, the number of free-floating brown dwarfs, formed from core collapse (‘like stars’) is expected to be significantly lower, only about 0.25\% of the number of stars, or 1-7 for the clusters considered here. Below 10$\,M_{\mathrm{Jup}}$ that number drops further by an order of magnitude. We also show that the planned surveys are not at risk of being significantly contaminated by field brown dwarfs in the foreground or background, after spectroscopic confirmation. Taken together, our results imply that if a population of L and T dwarfs were to be found in these JWST surveys, it is expected to be predominantly made up of rogue planets.
\end{abstract}

\section{Introduction}
\label{sec1}

Free-floating (or rogue) planets are objects that are formed in a disk around a young star, and subsequently ejected from their planetary system. According to simulations, these rogue planets are expected to be a significant component of the substellar population, in particular for masses below the Deuterium burning limit \citep[e.g.][]{vanelteren19,parker12}. So far we have very limited knowledge of this population. Deep surveys of clusters, star-forming regions, and the field have only been able to dip into the planetary-mass domain \citep[e.g.][]{scholz12,pena12,gagne17,lodieu21,miretroig22,bouy22}, and found a few dozen of spectroscopically confirmed objects below the Deuterium burning limit, which could have formed either from core collapse (‘like stars’) or in a protoplanetary disk (‘like planets’). Meanwhile, microlensing surveys tentatively indicate the existence of a population of free-floating objects with planetary masses \citep{mroz17,mcdonald21}. Our knowledge of the demographics of exoplanets will not be complete without firm observational constraints on the number of those that got ejected (or liberated) from their host planetary systems.

\begin{table*}[t]
\centering
    \caption{JWST GTO programs approved for cycle 1, with the potential to find free-floating planets.
    \label{tab:surveys}}
    \begin{tabular}{lcccc}
    \noalign{\smallskip}
    \tableline
    \noalign{\smallskip}
    Target            & NGC1333 & IC348 & NGC2024 & ONC \\
    \noalign{\smallskip}
    \tableline
    \noalign{\smallskip}
    Program ID        & 1202 & 1229$^a$ & 1190 & 1256 \\
    Instrument        & NIRISS & NIRCam & NIRCam & NIRCam \\
    Mode              & Slitless spectroscopy & Imaging & Imaging & Imaging \\
    FOV (sqarcmin)    & 30 & 20 & 10 & 80 \\
    On-source time (s) & 3100 & 300 & 6660 & 770\\
    Depth (K-mag)     & 21 & 23 & 24 & 23 \\
    Distance (pc)$^b$ & 296 & 324 & 414 & 403 \\
    Number of stars$^c$   & 200 & 500 & 800& 2600 \\
    Scheduled for$^d$    & 8-10/2023 & 8-10/2022 & 2-3/2023 & 9-10/2022\\
    \noalign{\smallskip}
    \tableline
    \noalign{\smallskip}
    \end{tabular}
    
$^a$ Program 1229 also includes a NIRSpec spectroscopy run at a later time, and parallel observations with NIRISS.
$^b$ From \citet{kuhn19}, except for NGC2024 where we use the distance cited in \citet{vanterwisga20}.
$^c$ From \citet{luhman16} for Perseus and \citet{kuhn15} for Orion.
$^d$ The JWST schedule is dynamic, that means, the actual observing dates may be different.
\end{table*}

The advent of JWST offers a tremendous boost to this field of research. It is the first telescope that has the capability to find and characterise directly free-floating planets down to masses of approximately 1$\,M_{\mathrm{Jup}}$ or even below. For the first time, we will be able to produce a robust census of the sub-stellar population with masses between 1 and 15$\,M_{\mathrm{Jup}}$, and probe directly the predictions from the simulations mentioned above. In cycle 1, four guaranteed time observing (GTO) programs will be in position to do exactly that, targeting four different young clusters of stars, two in Perseus and two in Orion, all within 500\,pc. Beyond cycle 1 a wide range of programs to survey other regions and to characterise the first samples of free-floating planets can be expected.

Any survey that aims to find free-floating planets in a specific star forming region will also find objects that are forming like stars in the same region, as well as contaminating brown dwarfs in the foreground and background of the cluster that share spectral characteristics with free-floating planets. In this paper, we aim to gauge likely outcomes of  the planned JWST programs, by estimating how many free-floating planets we expect to find in a given region, and how that number compares to the total number of objects with similar characteristics that will be detected in the same surveys. 

In the following we will distinguish between brown dwarfs, as objects that are formed alongside stars in the collapse and fragmentation of molecular clouds, and free-floating or rogue planets, as objects formed in and then ejected from a (proto-)planetary system. We acknowledge that at the time of writing there is no observational test available to distinguish between formation scenarios for individual objects, although research on exactly that topic will also benefit greatly from JWST observations. Most of the observational evidence to date supports the view that free-floating substellar objects down to the Deuterium burning limit are formed predominantly ‘like stars’ \citep[see review by][and references therein]{luhman12}. It is also plausible to assume that this population will diminish in number per mass bin as we probe lower and lower masses and approach the opacity limit for fragmentation \citep{bate12}. The purpose of this paper is to find out what type of objects, from which formation channel, we should expect to find below the Deuterium burning limit, based on our current understanding. Therefore it makes sense to distinguish populations based on formation scenarios. 

We proceed as follows. In section \ref{sec2} we summarise the GTO programs aiming to find free-floating planets in young clusters. In section \ref{sec3} we collate and contextualise predictions from work on simulating the dynamical evolution of planetary systems, to estimate the number of free-floating giant planets in young clusters. In section \ref{sec4}, we estimate the expected number of brown dwarfs (i.e. formed ‘like stars’) in the planetary mass domain. We devise a method to estimate the number of field brown dwarfs that may contaminate surveys for free-floating planets in section \ref{sec5}. Finally, we summarize and discuss these estimates in section \ref{sec6}.

\section{Summary of relevant JWST cycle 1 programs}
\label{sec2}

In JWST’s cycle 1, four GTO programs will conduct deep surveys in nearby star forming regions, and are in principle able to find young free-floating objects down to masses comparable to Jupiter. The four target regions are all within 500\,pc, two in Perseus (NGC1333 and IC348) and two in Orion (NGC2024 and the Orion Nebula Cluster, short ONC). These four programs are summarised in Table \ref{tab:surveys}, with their basic parameters estimated from the published survey design. While the survey in NGC1333 with the WFSS mode of the instrument NIRISS \citep{willott22} is a spectroscopy campaign, the remaining three are designed as multi-band imaging surveys with NIRCam, likely to be followed by spectroscopy.\footnote{We note that another program with similar science goals is GO2640, targeting the distant (6\,kpc) cluster Westerlund 2 with NIRCam.}

The four clusters offer a diverse set of initial conditions. NGC1333 and IC348 are of moderate density, host a few hundred stars and brown dwarfs each, but are without any O stars \citep{luhman16}. NGC2024 and the ONC are rich clusters, embedded in HII regions heated by massive stars (although in the case of NGC2024 the source of the heating is unclear, see \citet{vanterwisga20} and references therein). NGC1333, NGC2024 and the ONC are very young, with typical ages below or around 1 Myr, while IC348 is a few Myr old. In all four clusters, deep surveys from the ground or with HST have revealed a rich substellar population \citep{levine06,scholz12,dario12,alves13}.

All four programs have been scheduled in the JWST Long Range Plan, with anticipated observations in late 2022 for IC348 and the ONC, and 2023 for NGC2024 and NGC1333. In Table \ref{tab:surveys} we estimate the survey footprint and the K-band depth for the observations. The footprint (the field of view) is easy to infer from the design of the mosaics as shown in the Astronomer’s Proposal Tool (APT). The depth can be estimated from the given on-source exposure times using the JWST Exposure Time Calculator (ETC, version 1.7.0.1). We adopt a substellar source with a $T_\mathrm{eff}=1000$\,K model spectrum, and change the normalization in the K-band until sufficient signal-to-noise ratio (SNR) is achieved. We assume a desired SNR of 10, reached in the wide- and medium band filters used in these four campaigns, with central wavelength $>1.5\,\mu m$. Images at shorter wavelength may not be sensitive to the same level. We use the detector setup as specified in the APT descriptor, and medium background for all regions.

The NIRISS observations in NGC1333 will  provide immediate spectral information in the H- and K-bands, with a slitless instrument, as a result these are relatively shallow, with a projected depth at $H\sim22$ or $K\sim21$ \citep{willott22} The three NIRCam imaging programs are deeper, by 2-3\,mag, but will require spectroscopy follow-up. The observations in the ONC cover the widest field, the ones in NGC2024 have the longest integration times (but for a small field). The choice of filters and wavelength coverage, not discussed here further, is also different from program to program. All programs have their individual strengths, and are highly complementary to each other.

How do the projected depths of the programs compare to the instrument sensitivities as determined from JWST commissioning data \citep{rigby22}? For NIRISS in WFSS mode, the throughput is 30\% better, resulting in on-sky sensitivities that are 7-20\% improved compared to the ETC. For NIRCam, the throughput is comparable to pre-launch expectations, and the point source sensitivity is slightly better: 7.3 compared to 10\,nJy, for a SNR of 10 in 10000\,s exposures. This estimated sensitivity corresponds to a K-band (Vega) magnitude of about $K_{\mathrm{lim}}=27.3$, compared to $27.0$ for the value taken from the ETC. 

Infrared images of star forming regions are challenging for source extraction and precise photometry, due to the strongly varying background. The early released images of NGC3324, program ID\,2731 \citep{pontoppidan22},  may provide a good example of what to expect from NIRCam imaging of star forming regions. The pipeline-provided photometry catalogue for the F200W image with a central wavelength of $2.0\,\mu m$ and an on-source exposure time of 1610\,s is complete up to a Vegamag around 25 in this band. The F200W filter has approximately the same zeropoint as the 2MASS K-band ($\pm 0.2$\,mag). Scaling to 10000\,s, this is still 1.3\,mag shallower than $K_{\mathrm{lim}}$ as given in the commissioning report. The ETC estimates a SNR of $\sim 10$ for the Carina imaging at the completeness limit achieved in the actual data. Thus, while imaging of star forming regions unsurprisingly is not going to be as sensitive as in typical fields, these cross-checks provide re-assurance that the numbers from the ETC are realistic, albeit perhaps slightly pessimistic.

All programs listed in Table 1 aim to measure the mass function, therefore it seems appropriate to give some indications about the depth in terms of mass. A useful benchmark is a $1\,M_{\mathrm{Jup}}$ mass object at the age of around 1\,Myr. According to ‘hot start’ models of planet formation, such an object should have a brightness of $M_K=14$ \citep{spiegel12}, which translates to $K=21.3$ for Perseus or $K=22$ for Orion, all for $A_V=0$. This is comparable to the predictions from Lyon isochrones, which give $M_K=13.7$ (COND)  or $M_K=14.6$ (DUSTY) for this type of object \citep{baraffe02}. 

All programs in Table 1 should reach this limit. In fact, the NIRISS survey in NGC1333 was designed explicitly for this mass and age limit. The three imaging surveys will be deeper, and depending on the depth of the follow-up spectroscopy could reach objects with masses below $1\,M_{\mathrm{Jup}}$. The spectroscopic survey for IC348, already planned for cycle 1, should approximately match the depth of the imaging campaign. In this paper we will focus on the mass range between 1 and 15 $\,M_{\mathrm{Jup}}$ for which all surveys are sensitive.

We note that when making statements about the depth of the surveys, age and extinction need to be specified. With $A_V=10$, the brightness would drop by 1\,mag in the K-band. Also, with increasing age the brightness drops, to $M_K=16-17$ for 5\,Myr.

As another note of caution, with ‘cold start’ models, the predicted brightness of planets would drop by several magnitudes, to $M_K=17-18$ \citep{spiegel12}. If these are realistic, the mass range of the JWST surveys would be severely limited (for example, in the case of NGC1333, the survey would only reach about $5\,M_{\mathrm{Jup}}$). Recent observational tests of these model tracks \citep{flagg19,berardo17} as well as theoretical considerations \citep{marleau14,marleau19} make hot start scenarios far more plausible, hence, we ignore the cold start scenario in the following.

\section{Simulations of planet ejection}
\label{sec3}

Several groups have simulated the dynamical fate of planetary systems in clustered birth environments \citep[see][for a review]{parker20}. These simulations, with a varied set of initial conditions and assumptions, generally predict the presence of a population of free-floating planets in star forming regions. The main mechanisms that are responsible for the liberation of young planets from their host stars are encounters with other stars in a cluster and planet-planet interactions. Most of the cited work below comes from N-body simulations, using various approaches to reduce the complexity of the numerical problem. 

The predicted fraction of ejected planets in clusters from these simulations typically ranges from about 5 to 25\% \citep[e.g.,][]{parker12,liu13,zheng15,forgan15}. This rate depends on a wide variety of parameters. It increases with higher cluster density and substructure, and could perhaps reach 50\% or more for very dense, structured clusters \citep{daffern22} or to 1-2\% for low-density clusters \citep{cai17}. For environments like NGC1333 with moderate initial density (500-2000\,$M_{\odot}$pc$^{-3}$, \citealt{parker17}), a realistic estimate is probably at the lower end of the quoted range. In fact, \citet{adams06} argued that for a cluster like NGC1333 the disruption of planetary systems must be rare. For massive, dense clusters like the ONC, we would expect higher ejection rates (14\% in the simulations by \citealt{vanelteren19}, which mimic the ONC). On the other hand, disks may be disrupted more often in high-density environments, either by dynamical encounters with stars or by the radiation field from massive stars \citep{winter22}, inhibiting planet formation, and thus reducing the number of free-floating planets in an indirect way. The ejection rate may also depend on the properties of the host stars, in particular, it may increase with stellar mass \citep{fujii19}. Binary systems are more likely to eject planets than single stars, thus, the binary fraction in a cluster matters as well \citep{wang20}.

The age of the cluster is critical: While the number of ejected planets in a cluster rises strongly in the first few Myrs after planet formation, and then plateaus, a significant portion of the liberated planets will escape from the cluster, also on timescales of a few Myrs \citep{vanelteren19}. This means that free-floating planets are only expected to be found in young clusters for a relatively narrow age range, conceivably between 1 and 5\,Myr. For extremely young clusters, planets are not formed yet or have not had time to experience events that may lead to ejection. For slightly older clusters, the free-floating planets will have escaped into the field population (see also \citealt{smith01}). As noted in Section \ref{sec2}, the brightness of these objects drops steeply with age, making them much harder to detect in older clusters. 

So far, simulations have not explored the combination of all these effects systematically. Taken together, we would expect significant variations in the number of free-floating planets as a function of environmental parameters, such as the space density of stars. However, it may be challenging to directly measure these variations in the mass function, especially given the low numbers that are expected; see the discussion in \citet{muzic19}. We note that during main-sequence and post-main-sequence lifetimes more planets will be liberated \citep{verasmoeckel12,veras11}, which would increase the number of free-floating planets in the field, but not affect star forming regions. Observations with JWST, no matter the outcome, will put strong constraints on the quoted simulations. 

In the following we use the simulations as guidance to estimate the number of expected free-floating planets in young clusters with masses above 1$\,M_{\mathrm{Jup}}$, which are directly detectable with JWST (see Section \ref{sec2}). Given the dependencies outlined above, this is aimed to be an order of magnitude estimate.

To start, we need to establish how many giant planets are likely to be formed in the first place. Direct imaging surveys for giant planets typically detect planets with masses $>1\,M_{\mathrm{Jup}}$ and orbits $>5$\,au around a few percent of their target stars \citep{biller13,vigan17,gaudi21}. Specifically, for young systems \citet{vigan21} find that $\sim 6$\% host at least one such planet. In addition, long term radial velocity studies constrain the occurrence rate for $>1\,M_{\mathrm{Jup}}$ planets on orbital separations larger than 1\,au to about 5-15\% \citep{wolthoff22,wittenmyer20,fulton21}. About 1-3\% of stars will host a hot or warm Jupiter on orbits $<1$\,au \citep{wright12,wittenmyer20}. Microlensing studies indicate that 17\% of stars have planets of 0.3-10$\,M_{\mathrm{Jup}}$ and separations 0.5-10\,au \citep{cassan12}.

Note that all these studies focus on stars older than the star forming regions we are interested in. They will therefore not include planets that have been ejected or planets that have plunged into the star during the first Myrs. Thus, the occurrence rate may be significantly higher for young stars. Overall, it is plausible to assume that at least around 20\% of very young stars form a planet with a Jupiter mass or more, most of these will reside on separations of 1\,au or larger. With the ejection rates from simulations, as quoted above, that means between 1 and 5\% of stars eject a giant planet.

The approximate number of young stars in the four clusters in question is cited in Table 1. For NGC1333 we would expect about 40 of them to have formed a giant planet, for IC348 100, for NGC2024 160 and for the ONC 520. Again using the ejection rates from simulations, the number of free-floating planets with masses above 1$\,M_{\mathrm{Jup}}$ would be 2-10 in NGC1333, 5-25 in IC348, 8-40 in NGC2024, and 26-130 in the ONC. These numbers are summarised in Table \ref{tab:mf}.

\begin{figure}[t]
\center
\includegraphics[width=1.0\columnwidth]{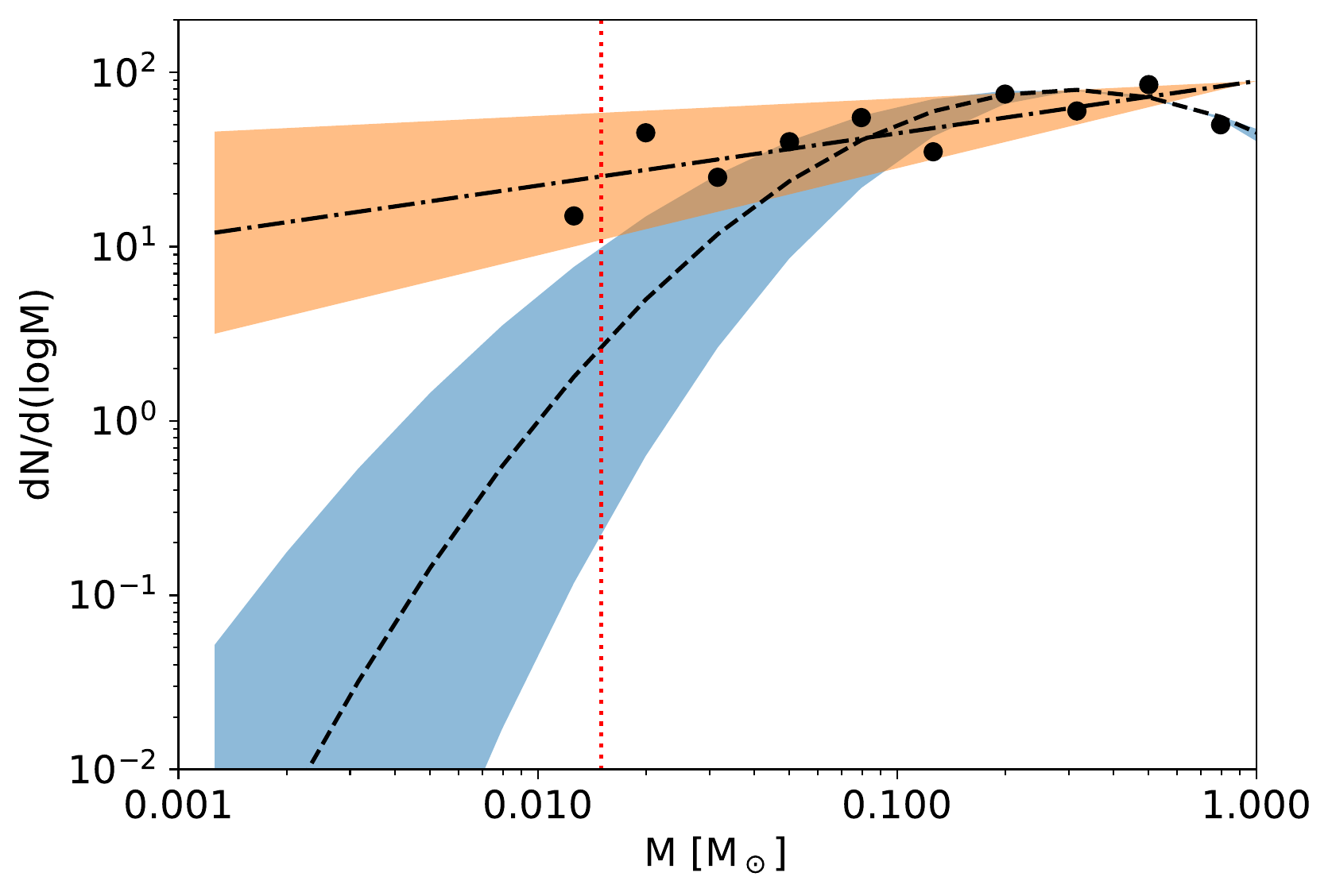}
\caption{Mass function in NGC1333 (black dots), matched by a log-normal mass function (dashed line) with $M_c=0.3\,M_{\odot}$ and $\sigma=0.5$, and a power-law mass function (dash-dotted line) with $\alpha=0.7$. The shaded region show the variation when the parameters of the mass functions are changed in the plausible range (see text). The red dotted line marks the upper mass limit considered in our estimates. 
\label{mf_fig}}
\end{figure}

The estimates come with a number of serious caveats. Just to name a few: As mentioned above, some of these rogue planets will leave the cluster on timescales comparable to the age of the stars, and some may not have had sufficient time yet to get ejected. Also, the surveys introduced in Section 2 do not cover the entirety of the clusters. Thus, not all these planets will reside in the cluster areas covered by JWST in cycle 1. Follow-up observations to cover wider areas to complete the census may be warranted.

The entire architecture of the young planetary systems has not been fully included yet in the available simulations of ejections. The ejection rate may be a function of planet orbital separations, planet masses, and overall numbers of planets. Inner planets may be better protected against ejection \citep{cai17}, which may significantly change the numbers and characteristics of rogue planets. At least for separations $>5$\,au, the simulations by \citet{vanelteren19} do not show such dependencies in single and multi-planet systems.

Maybe most importantly, most of the cluster simulations quoted above focus on the disruptions caused by stellar encounters. The impact of processes internal to the young planetary system, like planet-planet interactions and planet-disc interactions \citep{baruteau14}, on the free-floating population has not been fully explored yet. Planet-planet interactions in particular will cause instabilities, drive up eccentricities, and lead to ejections \citep{carrera19,li21}. Planet-planet interactions will especially affect giant planets; they can act as 'catalysts for planetary system disruption' \citep{cai17}. So far, however, in simulations attempting to quantify the population of free-floating planets, the outcomes are largely determined by encounters with stars, partly due to the simplified setup of planetary systems. For example, in the simulations by \citet{vanelteren19} the majority of the ejected planets have experienced an encounter with another star within 1\,Myr prior to ejection (about 70\%). 

If stars typically produce several giant planets, from which most are ejected by planet-planet interactions, the free-floating planet population could be significantly larger than estimated above \citep{verasraymond12,li15}. \citet{emsenhuber21} discusses giant planet formation as a ‘self-limiting process’, meaning that more giant planets are likely to lead to more dynamical instabilities within the system, with ejections as one possible outcome. How many giant planets are typically formed at birth is still an open question. Internal processes will produce free-floating planets irrespective of cluster environment, therefore, the surveys of low or moderate density clusters (NGC1333 and IC348) are particularly relevant to constrain those mechanisms.

\begin{table*}[th]
\centering
\caption{Number of very low mass objects according to a log-normal or a power-law mass function. 
\label{tab:mf}}
\begin{tabular}{llcccc}
\noalign{\smallskip}
\tableline
\noalign{\smallskip}
 & Fraction$^a$ & NGC1333 & IC348 & NGC2024 & ONC \\
\noalign{\smallskip}
\tableline
\noalign{\smallskip}
Free-floating planets$^b$               & 1-5\%     & 2-10 & 5-25   & 8-40   & 26-130\\
Log-normal MF, 1-15$\,M_{\mathrm{Jup}}$ & 0.25\%    & 1    & 1      & 2      & 7\\
-- range                                & 0.1-1.4\% & 0-3  & 0-7    & 0-11   & 0-36 \\
Power-law MF, 1-15$\,M_{\mathrm{Jup}}$  & 11\%      & 22   & 55     & 88     & 286\\
-- range                                & 4-31\%    & 8-62 & 20-155 & 32-248 & 104-806\\
Contamination L/T dwarfs$^c$            &           & 1.4  & 15     & 0.2    & 1.0\\
\noalign{\smallskip}
\tableline
\end{tabular}

$^a$ Fraction of objects relative to the total number of stars and brown dwarfs in a cluster (see Table \ref{tab:surveys}).
$^b$ According to the estimates in Section \ref{sec3}.
$^c$ According to the estimates in Section \ref{sec5}.
\end{table*}

\section{Planetary-mass brown dwarfs}
\label{sec4}

In addition to ejected planets, we expect that young clusters will harbor brown dwarfs with masses comparable to giant planets that are formed by cloud fragmentation and core collapse, i.e. in a way similar to stars. Here we estimate how many such objects we expect to find in deep surveys. 

 Star formation simulations predict that the outcome is a population of stars with masses in agreement with a log-normal mass function \citep{bate12}, with a cutoff at the opacity limit for fragmentation. A commonly quoted explanation for this is the central limit theorem \citep{offner14}. The lower mass limit has not been determined yet empirically, but is expected to be in the range between 1 and 10$\,M_{\mathrm{Jup}}$ \citep{bate12}. The statistics of the stellar population predicted by simulations  will depend on the specific formulation and ingredients, e.g., turbulence \citep{padoan04}, radiative feedback \citep{guszejnov16}, stellar winds \citep{krumholz12}, magnetic fields \citep{price07,wurster19} are all expected to alter the shape of the IMF and the star formation rate significantly.

Observationally, however, the log-normal mass function for local galactic star forming regions is found to be universal, or near universal \citep{bastian10}. The parameters of the log-normal distribution, the critical mass $M_c$ and the width $\sigma$, have been empirically robustly determined for a wide range of clusters. \citet{damian21} measured the mass function in 8 clusters with a range of environmental parameters, and find on average $M_c=0.32\pm0.02$ and $\sigma=0.47\pm0.02$. All observational surveys to date of galactic young clusters are consistent with $M_c=0.25-0.35$ and $\sigma=0.4-0.6$. This is the range of parameters we adopt here. 
Instead of the log-normal form, many observational surveys parameterise the mass function as a power law $dN/dM \sim M^{-\alpha}$. Observed mass functions for brown dwarfs are consistent with $\alpha=0.7$, with values ranging from 0.5 to 0.9, for many clusters \citep{muzic17} and also in the field \citep{kirkpatrick21}. Again this points to a universal mass function. In the following we therefore assume that all clusters share a consistent underlying mass function for the objects that form from core collapse.

To determine the empirical scaling of the mass function, we use the sample of 100 members of the cluster NGC1333 with estimated masses from \citet{scholz12}, covering the domain of brown dwarfs and low-mass stars. The same sample has also been used in \citet{offner14} and is shown to be consistent with the low-mass mass function in many other clusters. The sample comprises about half of the known members of NGC1333. Compared to the census by \citet{luhman16}, which comprises about 200 members, our IMF sample is missing the stars around and above $0.7\,M_{\odot}$ as well as some objects in the outskirts of the cluster. 

The mass function for NGC1333 together with a log-normal parameterisation is shown in Fig. \ref{mf_fig}. The log-normal mass function has been shifted to match the datapoints for high-mass brown dwarfs and very low mass stars. We also show in Figure \ref{mf_fig} the standard power-law mass function, with a slope of $\alpha=0.7$ (for $dN/dM$), or $\Gamma=-0.3$ (for $dN/dlog(M)$). This has again been shifted to match the datapoints in NGC1333. For about an order of magnitude in masses, from 0.05 to 0.5\,$M_{\odot}$, log-normal and power-law parameterisation are in good agreement. They diverge strongly in the planetary mass domain below the Deuterium burning limit.

We note that for the two lowest mass bins in NGC1333 the current data clearly exceed the expected log-normal mass function. A similar excess has been observed in a number of other clusters and associations, \citep{miretroig22,gagne17,muzic19}, see also Figure 2 in \citet{offner14}. This could be an indication of a second population, potentially ejected giant planets, but at this stage the data are not sufficiently robust to be confident.

We can now add up the expected number of objects according to the assumed underlying mass function, at first for NGC1333. To do that we use the mass functions shown in Figure \ref{mf_fig}. With a log-normal mass function, as expected for objects forming like stars, we expect 0.5 (between 0 and 3) objects with masses between 1 and 15$\,M_{\mathrm{Jup}}$ -- but since the mass function is declining steeply, most of these would be near the top end of the relevant mass range ($M>10M_{\mathrm{Jup}}$). This corresponds to 0.25\% of the known cluster population in NGC1333, including stars and brown dwarfs. The power-law mass function on the other hand predicts 21 (7-62) objects in the same mass range, or 11\% of the cluster population. More than half of these have masses below $10\,M_{\mathrm{Jup}}$.

These results can easily be scaled to other clusters, again assuming the universality of the mass function, and using the number of stars and brown dwarfs in the cluster as scaling factor. In Table \ref{tab:mf} we summarise these estimates for the four clusters in question here. 

\begin{figure*}[t]
\center
\includegraphics[width=1.0\columnwidth]{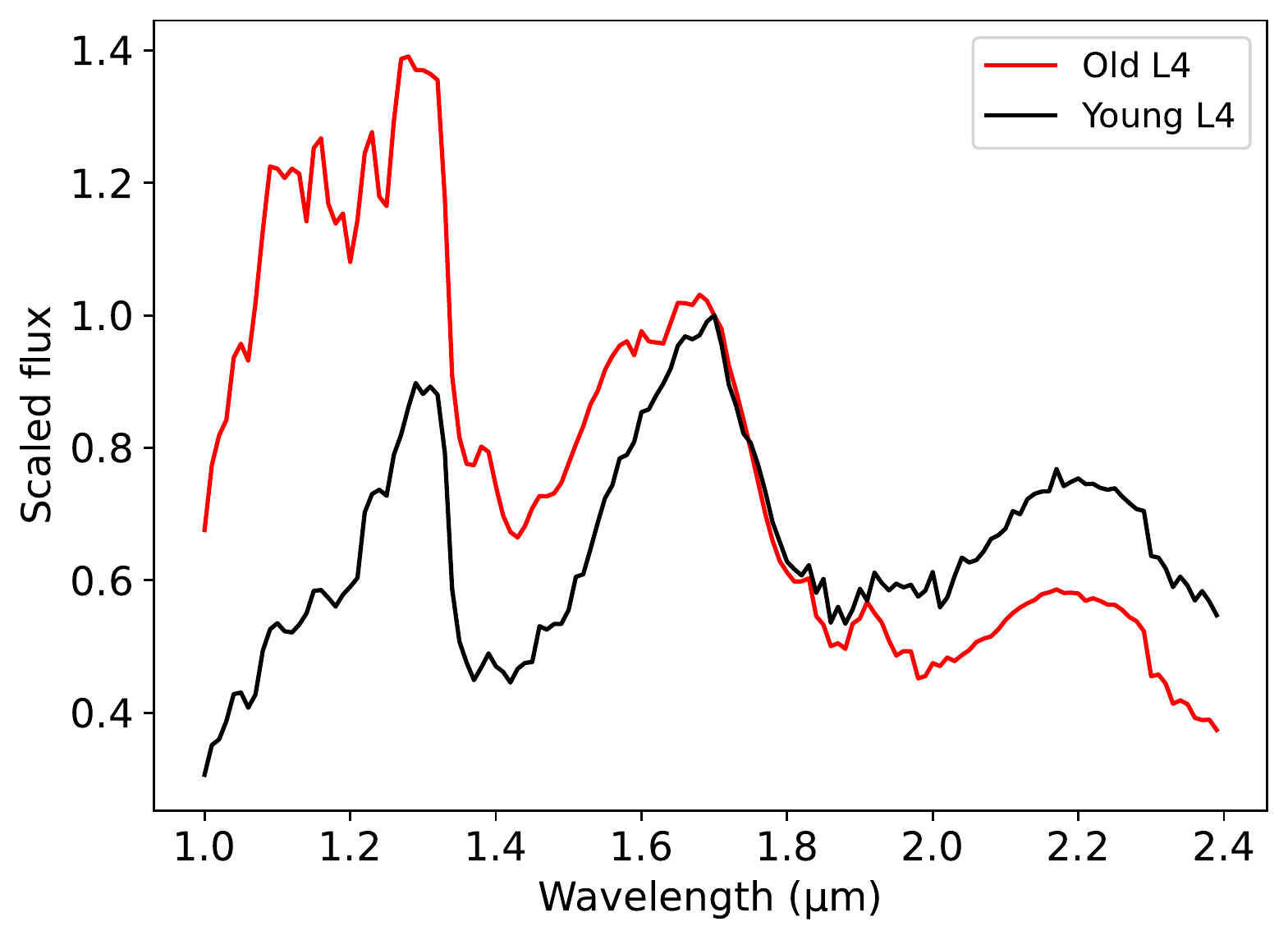}
\includegraphics[width=1.0\columnwidth]{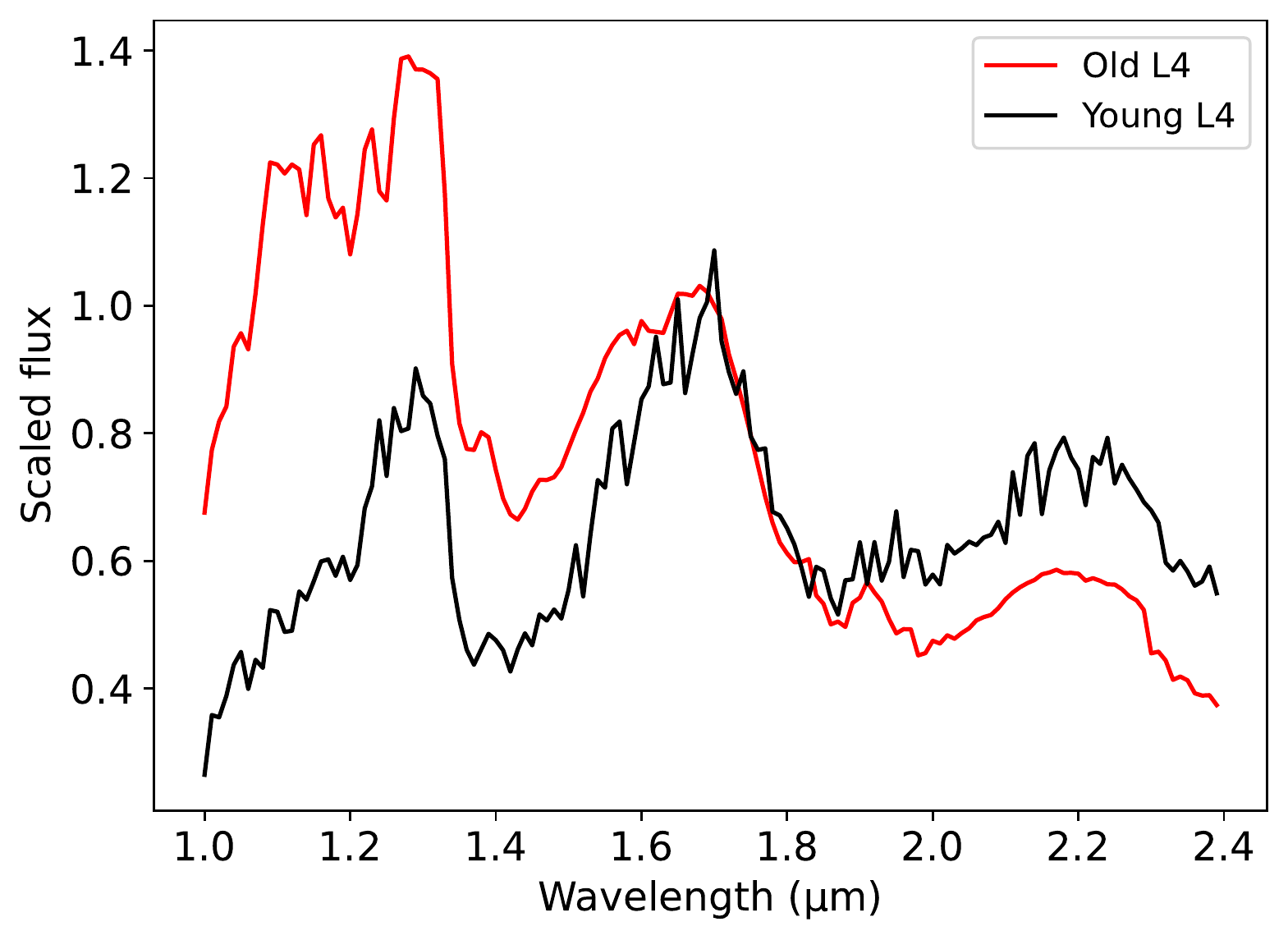}
\caption{Comparison between spectra for young and old brown dwarfs, both at spectral type L4. The young template is from \citet{luhman17}, the old is DENIS-P J170548.38-051645.7, published as template in \citet{burgasser10}, obtained from the SpeX Prism Spectral Libraries. All spectra are rebinned to a resolution of 0.01$\,\mu m$ using {\tt Spectres} \citep{carnall17}, and scaled to an arbitrary flux level. On the left side, we show both template spectra at high SNR, on the right side we added Gaussian noise (for a SNR of 20) to the young template.
\label{spectrum_fig}}
\end{figure*}

This exercise demonstrates that the population of objects below the Deuterium burning limit that are formed ‘like stars’ from core collapse, and form the continuation of the log-normal mass function, is expected to be very small. In particular, it will be negligible for masses of $1-10\,M_{\mathrm{Jup}}$. The numbers of expected objects from core collapse is far below the expected numbers of ejected planets, especially in the 1-10\,$M_{\mathrm{Jup}}$ mass domain. In contrast, the power-law mass function predicts a substantial population of substellar objects down to masses comparable to Jupiter, exceeding the estimates for ejected planets we made in Section \ref{sec3}.

The analysis shows how valuable the JWST surveys will be compared to existing ground-based observations: The current surveys are typically incomplete below $10\,M_{\mathrm{Jup}}$. Given the uncertainties it is difficult to distinguish between an extended power-law and a log-normal mass function. On the other hand, JWST can probe sufficiently deep to make that distinction. If the observed mass function of free-floating objects below the Deuterium burning limit is found to be inconsistent with the log-normal mass function, we would have clear evidence for the presence of free-floating, ejected planets in the observed clusters. 

\section{Contamination by field brown dwarfs}
\label{sec5}

Every cluster survey is sensitive to a range of objects in the foreground and background of the target region. In surveys for young L and T dwarfs, the most problematic type of contamination is from field brown dwarfs of the same spectral type. While young L and T dwarfs have characteristic spectra, as well as different colours \citep{cruz09,almendros22}, these distinctions can be difficult to detect at low signal-to-noise ratios. In Figure \ref{spectrum_fig}, we show a comparison of spectra for a young and an old template, for spectral type L4, for a resolution comparable to what can be achieved with NIRISS/WFSS or NIRSpec/PRISM. For the young template, we also add a version with 5\% noise of the signal (or SNR of 20). The differences in the shape of the spectrum as well as the overall relative fluxes in different bands can still be appreciated at this noise level. If more noise is present, the differences between young and old become difficult to discern.

Additionally, in the absence of spectra, more exotic objects in the background, particularly very late giants and red-shifted AGNs, could in principle mimic the colour signature of young brown dwarfs. We will neglect those exotic background contaminants here because we assume that any substellar candidate will need a spectrum (or a complete spectral energy distribution in the near/mid-infrared) for confirmation (see also \citealt{almendros22} for further discussion). We note that in regions with strong extinction (ONC, NGC2024), the cloud acts as a screen and prevents background contamination that way. 

This leaves field brown dwarfs along the pencil beam towards the cluster as the most important source of potential contamination. To estimate the number of expected contaminating objects in a survey, we need to know the survey depth in a given band, as well as the survey footprint. Those numbers are summarised in Table \ref{tab:surveys} for the four cycle 1 projects investigated here, obtained from the project description as currently published (see Section \ref{sec2}). In addition, we obtain the space densities as a function of spectral type from \citet{best21}. We use the parametric form of the absolute J-band magnitude as a function of spectral type provided by \citet{filippazzo15}, and convert to absolute K-band magnitude using the (J-K) colour vs. spectral type relation \citep[Figure 6 in][]{best21}.

We choose here to divide the relevant spectral range into four bins, L0-L5, L5-T0, T0-T5, and T5-T9. The space densities for those bins are 2.22, 2.20, 1.12, 4.48\,$\times 10^{-3}$ pc$^{-3}$, with an error of 5-10\% \citep{best21}. The absolute magnitudes in the K-band are estimated to be 11.3, 13.0, 14.4, 16.9, with a considerable spread of about 0.5 mag. With those absolute magnitudes, and the survey depth (an apparent magnitude), we can estimate the limiting distance of the survey as a function of spectral type. For NGC2024 and the ONC we adopt the cluster distance as maximum limiting distance, due to the screening effect of the cloud. With the survey footprint, we can then define a survey volume for each project -- a pyramid extending to the limiting distance. Multiplying this volume with the space densities yields the expected rate of contamination. 

\begin{figure}[t]
\center
\includegraphics[width=1.0\columnwidth]{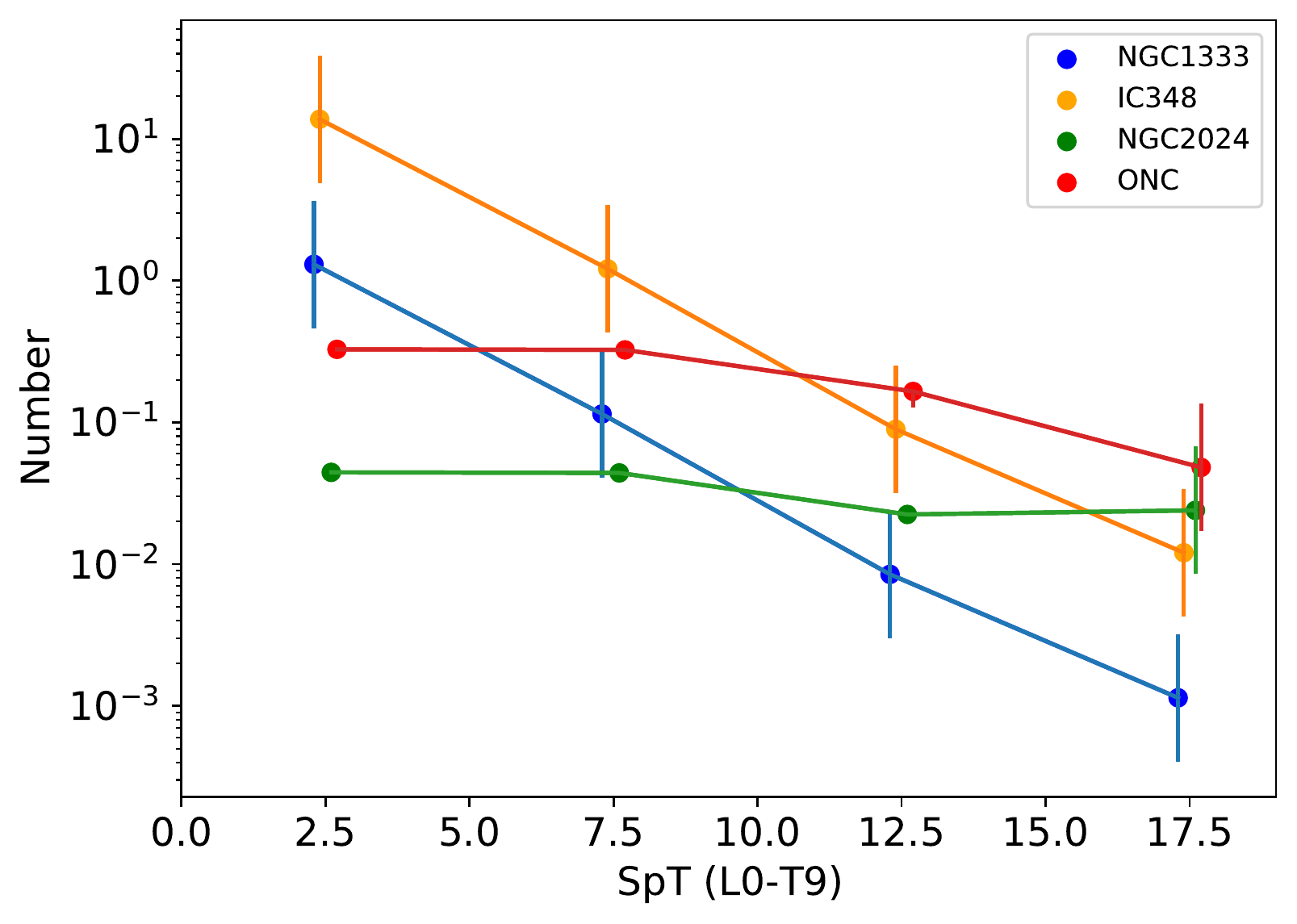}
\caption{Number of contaminating field dwarfs in the four surveys discussed in the text, as a function of spectral types. Spectral types are numerically defined with L0 being 0.0 and T9 being 19.0. Errorbars take into account $\pm$0.5\,mag uncertainty in survey depth and $\pm$0.25\,mag error in absolute magnitudes. For NGC2024 and the ONC, we assumed that the cloud acts a screen to eliminate background contamination. 
\label{contam_fig}}
\end{figure}

Results from this exercise are summarised in Figure \ref{contam_fig}, for the four surveys and as a function of spectral type. The errorbars in this figure take into account $\pm 0.5$\,mag uncertainty in survey depth, as well as $\pm 0.25$\,mag in the absolute magnitudes. All other uncertainties are negligible. For NGC1333, we expect little contamination by field dwarfs -- in total, the calculation yields 1.4 objects in this field (with a range 0.5-4.0), most of them are going to be early L dwarfs. IC348 could in principle find considerable numbers of contaminating brown dwarfs. Our calculation yields 15, with a range from 5 to 42.  Similar to NGC1333, most are expected to be early L dwarfs. For the ONC the total number is 0.9, with a relatively small errorbar (0.8-1.0), and almost uniformly distributed with spectral type. This is a result of defining the cloud as a screen that eliminates background contamination, and thus a portion of the bright L dwarfs. For NGC2024 the expected numbers are very small, in total less than 0.2, due to the small field of view. Again the contamination rate is flat vs. spectral type due to the screening of the cloud. We note that qualitatively similar results were obtained by \citet{caballero08} using a different method and a different test case.

These numbers should be considered upper limits, for a variety of reasons.  For one, most of the contamination is expected for early L dwarfs, which are visible to larger distances. Since young early L dwarfs are bright compared to the survey depth, distinguishing them from the field brown dwarfs with spectroscopy should be straightforward (see Figure \ref{spectrum_fig}). The WFSS survey in NGC1333 will immediately give spectral information, and the one in IC348 is associated with a follow-up spectroscopy campaign with NIRSpec. Similar campaigns can be expected for the Orion observations. 

Second, with spectroscopy included, the surveys will not reach the depth expected for imaging. There will be a number of faint objects at the limits of the imaging surveys for which spectroscopic follow-up is prohibitively expensive in terms of integration time. Adopting a putative spectroscopic depth of $K=22$, the number of contaminants drops to 3.8 for IC348, 0.8 for the ONC, and 0.1 for NGC2024. As a reminder, the estimated absolute K-band magnitude for a $1\,M_{\mathrm{Jup}}$, 1\,Myr free-floating planet is $M_K\sim 14$, which translates to $K\sim 21$ for Perseus and $K\sim 22$ for Orion (at $A_V=0$). Finally, while the extinction in NGC1333 and IC348 is moderate and will not completely block the background contamination, it will screen out some fainter sources and reduce the overall survey volume.

We also note that extinction is going to alter the spectral energy distributions for most young free-floating planets, which should help distinguishing them from reddening-free field brown dwarfs. Taken all that into account, contamination by field brown dwarfs is not going to be an issue for these surveys. However, spectroscopy with decent signal-to-noise ratio is going to be needed to confirm individual objects. 

While the space densities of brown dwarfs are only known for the local volume out to distances of 25\,pc, we assume here that they are broadly the same when probing larger distances in the foreground of the clusters investigated here. This is a plausible assumption: Recent work based on Gaia data does not show strong variations in stellar density out to distances of 500\,pc \citep{miyachi19,widmark22} towards the galactic directions of Perseus and Orion, relevant to this work. Similarly, the number counts of stars from the Besancon Galaxy model \citep{robin03} only change within 10\% of the total as a function of galactic sightline for the regions investigated here. If anything, the number densities are expected to drop off for distances $>100\,$pc, due to the galactic structure \citep[see Fig. 1 in][]{caballero08}, again confirming that our estimates should be seen as upper limits. 

\section{Summary and discussion}
\label{sec6}

In the preceding sections we have investigated the expected numbers of free-floating planets in JWST surveys, plus the number of brown dwarfs formed by core collapse, plus the expected number of brown dwarf contaminants. Our estimates are tailored to four specific JWST cycle 1 programs, but can be translated and scaled easily to other target clusters in future programs. In this section, we first summarise our findings, and then put them into context.

In Section \ref{sec3} we showed that based on our current understanding and judged by the outcomes of existing simulations, we expect that nearby star forming clusters within 500\,pc host between a few to a hundred free-floating (or rogue) planets that are formed in a disk and are later ejected, with masses above $1\,M_{\mathrm{Jup}}$. These objects would be ejected either due to encounters with other stars or due to planet-planet interactions. This corresponds to a few percent of the number of stars in the cluster. These estimates come with a number of caveats. In particular, the numbers could be significantly higher if stars form typically several giant planets from which most are ejected. Empirical measurements will be key in guiding the development of simulations.

We also show, in Section \ref{sec4}, that the number of planetary-mass brown dwarfs that are likely to be formed by core collapse like stars, is very low, about 0.25\% of the number of stars for the mass range 1-15$\,M_{\mathrm{Jup}}$, and an order of magnitude lower for 1-10$\,M_{\mathrm{Jup}}$. That means, the mass function in this domain will be very different depending on which population dominates. If ejected planets are very rare and significantly less common than estimated above, we expect a steep dropoff following a log-normal mass function below the Deuterium burning limit. If ejected planets are as abundant as simulations predict, we should see a clear deviation from the log-normal mass function below $10\,M_{\mathrm{Jup}}$. Or put in simple terms: If observations find significant numbers of free-floating objects with masses comparable to Jupiter, those are expected to be rogue planets, and not very low mass brown dwarfs. 

Our estimates can be contrasted with existing observational programs that also target this mass range. In 2011, Sumi et al. published a contentious paper claiming that the number of planets that are unbound or on wide orbits exceeds the number of stars \citep{sumi11}. This claim triggered a lot of interest, but did not hold up to further scrutiny. \citet{mroz17} find a 95\% upper limit of 0.25 free-floating/wide orbit Jupiter-mass planets per star, estimated from microlensing detections, superseding earlier estimates from the same method. Some or most of these could be explained by wide (10\,au or wider) Jupiter-type planets \citep{clanton17}, i.e. the upper limit for free-floating planets would be smaller. Further constraints are expected from the Kepler/K2 microlensing campaign, but have not been published yet \citep{mcdonald21}. In the terminology of our paper, this estimate would include actual ejected planets and low-mass brown dwarfs formed by core collapse. Microlensing probes the field population; as pointed out in Section 3, we expect a larger number of free-floating planets per star in the field compared to young clusters, due to escape from the clusters and further ejections from planetary systems during stellar evolution. As more stars turn into white dwarfs or neutron stars, some of their planets are going to be released into the field population. Over timescales of Gyrs, the number of free-floating planets per star will therefore increase. Adding our estimates from Section \ref{sec3} and \ref{sec4}, we would expect in the range of $\sim 0.05$ free-floating object (planet or brown dwarf) with Jupiter- or super-Jupiter masses per star, consistent with the limit from microlensing. 

Direct searches for ultra-low mass free-floating objects in clusters are currently only sensitive to masses $>5\,M_{\mathrm{Jup}}$. These surveys tend to find that objects with masses below the Deuterium burning limit account for 1-14\% \citep{lucas06}, 2-5\% \citep{scholz12}, 2-7\% \citep{miretroig22}, and $\sim 10$\% \citep{pena12} of the number of stars and brown dwarfs in the studied clusters. We caution that determining masses for objects around the Deuterium burning limit is a difficult process, and therefore these numbers come with significant errorbars. As already indicated in Section \ref{sec4}, these existing estimates exceed what is expected from a log-normal mass function, as pointed out by several groups. Although these surveys are not as deep as the ongoing work with JWST, they are already finding numbers of planetary mass objects that are comparable to the sum of the numbers for free-floating planets and very low mass brown dwarfs estimated in this paper. This may indicate that ejected planets are more abundant than expected from the existing simulations, or that existing surveys overestimate the number of objects in this mass domain. Probing the mass ranges below $5\,M_{\mathrm{Jup}}$ with JWST will certainly provide clarification on this issue. 

Young rogue planets with Jupiter-like masses will have spectral types of L and T. In Section \ref{sec5} we also estimate the number of brown dwarfs in the foreground and background of the young clusters that may be found among those objects with L and T spectral types. For rich embedded clusters where the cloud shields the background, like the ONC, the contamination by brown dwarfs is orders of magnitude lower than the expected numbers of young objects residing in the clusters with those spectral types. For sparsely populated or low extinction environments, like NGC1333, the field brown dwarf contamination is still very low, and dominated by early L type objects which are straightforward to identify. Follow-up spectroscopy will be needed for a detailed characterisation, but even with low signal-to-noise ratio and resolution, the contamination by field brown dwarfs in these surveys will be negligible.

\section{Acknowledgements}

We would like to thank the referee, Richard Parker, for a concise and constructive report on our paper. 
AS, KM, and RJ are PI and Co-Is of the JWST/GTO program 1202, 'The NIRISS Survey for Young Brown Dwarfs and Rogue Planets'. KM acknowledges funding by the Science and Technology Foundation of Portugal (FCT), grants No. PTDC/FIS-AST/28731/2017 and UIDB/00099/2020. LQ contributed to this paper as part of a summer research project, for which funding was kindly provided by Rita Tojeiro and Vivienne Wild. This research has benefitted from the SpeX Prism Spectral Libraries, maintained by Adam Burgasser at {\url http://www.browndwarfs.org/spexprism}.


\begin{thebibliography}{}

\bibitem[Adams et al.(2006)]{adams06} Adams, F.~C., Proszkow, E.~M., Fatuzzo, M., et al.\ 2006, \apj, 641, 504. doi:10.1086/500393

\bibitem[Almendros-Abad et al.(2022)]{almendros22} Almendros-Abad, V., Mu{\v{z}}i{\'c}, K., Moitinho, A., et al.\ 2022, \aap, 657, A129. doi:10.1051/0004-6361/202142050


\bibitem[Alves de Oliveira et al.(2013)]{alves13} Alves de Oliveira, C., Moraux, E., Bouvier, J., et al.\ 2013, \aap, 549, A123. doi:10.1051/0004-6361/201220229

\bibitem[Baraffe et al.(2002)]{baraffe02} Baraffe, I., Chabrier, G., Allard, F., et al.\ 2002, \aap, 382, 563. doi:10.1051/0004-6361:20011638

\bibitem[Baruteau et al.(2014)]{baruteau14} Baruteau, C., Crida, A., Paardekooper, S.-J., et al.\ 2014, Protostars and Planets VI, 667. doi:10.2458/azu\_uapress\_9780816531240-ch029

\bibitem[Bastian et al.(2010)]{bastian10} Bastian, N., Covey, K.~R., \& Meyer, M.~R.\ 2010, \araa, 48, 339. doi:10.1146/annurev-astro-082708-101642

\bibitem[Bate(2012)]{bate12} Bate, M.~R.\ 2012, \mnras, 419, 3115. doi:10.1111/j.1365-2966.2011.19955.x

\bibitem[Beleznay \& Kunimoto(2022)]{beleznay22} Beleznay, M. \& Kunimoto, M.\ 2022, arXiv:2207.12522

\bibitem[Berardo \& Cumming(2017)]{berardo17} Berardo, D. \& Cumming, A.\ 2017, \apjl, 846, L17. doi:10.3847/2041-8213/aa81c0

\bibitem[Best et al.(2021)]{best21} Best, W.~M.~J., Liu, M.~C., Magnier, E.~A., et al.\ 2021, \aj, 161, 42. doi:10.3847/1538-3881/abc893

\bibitem[Biller et al.(2013)]{biller13} Biller, B.~A., Liu, M.~C., Wahhaj, Z., et al.\ 2013, \apj, 777, 160. doi:10.1088/0004-637X/777/2/160

\bibitem[Bouy et al.(2022)]{bouy22} Bouy, H., Tamura, M., Barrado, D., et al.\ 2022, arXiv:2206.00916

\bibitem[Burgasser et al.(2010)]{burgasser10} Burgasser, A.~J., Cruz, K.~L., Cushing, M., et al.\ 2010, \apj, 710, 1142. doi:10.1088/0004-637X/710/2/1142


\bibitem[Caballero et al.(2008)]{caballero08} Caballero, J.~A., Burgasser, A.~J., \& Klement, R.\ 2008, \aap, 488, 181. doi:10.1051/0004-6361:200809520

\bibitem[Cai et al.(2017)]{cai17} Cai, M.~X., Kouwenhoven, M.~B.~N., Portegies Zwart, S.~F., et al.\ 2017, \mnras, 470, 4337. doi:10.1093/mnras/stx1464

\bibitem[Carnall(2017)]{carnall17} Carnall, A.~C.\ 2017, arXiv:1705.05165

\bibitem[Carrera et al.(2019)]{carrera19} Carrera, D., Raymond, S.~N., \& Davies, M.~B.\ 2019, \aap, 629, L7. doi:10.1051/0004-6361/201935744

\bibitem[Cassan et al.(2012)]{cassan12} Cassan, A., Kubas, D., Beaulieu, J.-P., et al.\ 2012, \nat, 481, 167. doi:10.1038/nature10684

\bibitem[Clanton \& Gaudi(2017)]{clanton17} Clanton, C. \& Gaudi, B.~S.\ 2017, \apj, 834, 46. doi:10.3847/1538-4357/834/1/46

\bibitem[Cruz et al.(2009)]{cruz09} Cruz, K.~L., Kirkpatrick, J.~D., \& Burgasser, A.~J.\ 2009, \aj, 137, 3345. doi:10.1088/0004-6256/137/2/3345


\bibitem[Daffern-Powell et al.(2022)]{daffern22} Daffern-Powell, E.~C., Parker, R.~J., \& Quanz, S.~P.\ 2022, \mnras, 514, 920. doi:10.1093/mnras/stac1392

\bibitem[Damian et al.(2021)]{damian21} Damian, B., Jose, J., Samal, M.~R., et al.\ 2021, \mnras, 504, 2557. doi:10.1093/mnras/stab194

\bibitem[Da Rio et al.(2012)]{dario12} Da Rio, N., Robberto, M., Hillenbrand, L.~A., et al.\ 2012, \apj, 748, 14. doi:10.1088/0004-637X/748/1/14

\bibitem[Emsenhuber et al.(2021)]{emsenhuber21} Emsenhuber, A., Mordasini, C., Burn, R., et al.\ 2021, \aap, 656, A70. doi:10.1051/0004-6361/202038863

\bibitem[Filippazzo et al.(2015)]{filippazzo15} Filippazzo, J.~C., Rice, E.~L., Faherty, J., et al.\ 2015, \apj, 810, 158. doi:10.1088/0004-637X/810/2/158


\bibitem[Flagg et al.(2019)]{flagg19} Flagg, L., Johns-Krull, C.~M., Nofi, L., et al.\ 2019, \apjl, 878, L37. doi:10.3847/2041-8213/ab276d

\bibitem[Forgan et al.(2015)]{forgan15} Forgan, D., Parker, R.~J., \& Rice, K.\ 2015, \mnras, 447, 836. doi:10.1093/mnras/stu2504

\bibitem[Fulton et al.(2021)]{fulton21} Fulton, B.~J., Rosenthal, L.~J., Hirsch, L.~A., et al.\ 2021, \apjs, 255, 14. doi:10.3847/1538-4365/abfcc1

\bibitem[Fujii \& Hori(2019)]{fujii19} Fujii, M.~S. \& Hori, Y.\ 2019, \aap, 624, A110. doi:10.1051/0004-6361/201834677

\bibitem[Gagn{\'e} et al.(2017)]{gagne17} Gagn{\'e}, J., Faherty, J.~K., Mamajek, E.~E., et al.\ 2017, \apjs, 228, 18. doi:10.3847/1538-4365/228/2/18

\bibitem[Gaudi et al.(2021)]{gaudi21} Gaudi, B.~S., Meyer, M., \& Christiansen, J.\ 2021, ExoFrontiers; Big Questions in Exoplanetary Science, 2. doi:10.1088/2514-3433/abfa8fch2


\bibitem[Guszejnov et al.(2016)]{guszejnov16} Guszejnov, D., Krumholz, M.~R., \& Hopkins, P.~F.\ 2016, \mnras, 458, 673. doi:10.1093/mnras/stw315

\bibitem[Kirkpatrick et al.(2021)]{kirkpatrick21} Kirkpatrick, J.~D., Gelino, C.~R., Faherty, J.~K., et al.\ 2021, \apjs, 253, 7. doi:10.3847/1538-4365/abd107

\bibitem[Krumholz et al.(2012)]{krumholz12} Krumholz, M.~R., Klein, R.~I., \& McKee, C.~F.\ 2012, \apj, 754, 71. doi:10.1088/0004-637X/754/1/71

\bibitem[Kuhn et al.(2015)]{kuhn15} Kuhn, M.~A., Getman, K.~V., \& Feigelson, E.~D.\ 2015, \apj, 802, 60. doi:10.1088/0004-637X/802/1/60

\bibitem[Kuhn et al.(2019)]{kuhn19} Kuhn, M.~A., Hillenbrand, L.~A., Sills, A., et al.\ 2019, \apj, 870, 32. doi:10.3847/1538-4357/aaef8c

\bibitem[Levine et al.(2006)]{levine06} Levine, J.~L., Steinhauer, A., Elston, R.~J., et al.\ 2006, \apj, 646, 1215. doi:10.1086/504964

\bibitem[Li et al.(2015)]{li15} Li, Y., Kouwenhoven, M.~B.~N., Stamatellos, D., et al.\ 2015, \apj, 805, 116. doi:10.1088/0004-637X/805/2/116

\bibitem[Li et al.(2021)]{li21} Li, J., Lai, D., Anderson, K.~R., et al.\ 2021, \mnras, 501, 1621. doi:10.1093/mnras/staa3779

\bibitem[Liu et al.(2013)]{liu13} Liu, H.-G., Zhang, H., \& Zhou, J.-L.\ 2013, \apj, 772, 142. doi:10.1088/0004-637X/772/2/142

\bibitem[Lodieu et al.(2021)]{lodieu21} Lodieu, N., Hambly, N.~C., \& Cross, N.~J.~G.\ 2021, \mnras, 503, 2265. doi:10.1093/mnras/stab401

\bibitem[Lucas et al.(2006)]{lucas06} Lucas, P.~W., Weights, D.~J., Roche, P.~F., et al.\ 2006, \mnras, 373, L60. doi:10.1111/j.1745-3933.2006.00244.x


\bibitem[Luhman(2012)]{luhman12} Luhman, K.~L.\ 2012, \araa, 50, 65. doi:10.1146/annurev-astro-081811-125528

\bibitem[Luhman et al.(2016)]{luhman16} Luhman, K.~L., Esplin, T.~L., \& Loutrel, N.~P.\ 2016, \apj, 827, 52. doi:10.3847/0004-637X/827/1/52

\bibitem[Luhman et al.(2017)]{luhman17} Luhman, K.~L., Mamajek, E.~E., Shukla, S.~J., et al.\ 2017, \aj, 153, 46. doi:10.3847/1538-3881/153/1/46

\bibitem[Marleau \& Cumming(2014)]{marleau14} Marleau, G.-D. \& Cumming, A.\ 2014, \mnras, 437, 1378. doi:10.1093/mnras/stt1967


\bibitem[Marleau et al.(2019)]{marleau19} Marleau, G.-D., Mordasini, C., \& Kuiper, R.\ 2019, \apj, 881, 144. doi:10.3847/1538-4357/ab245b

\bibitem[McDonald et al.(2021)]{mcdonald21} McDonald, I., Kerins, E., Poleski, R., et al.\ 2021, \mnras, 505, 5584. doi:10.1093/mnras/stab1377

\bibitem[Miret-Roig et al.(2022)]{miretroig22} Miret-Roig, N., Bouy, H., Raymond, S.~N., et al.\ 2022, Nature Astronomy, 6, 89. doi:10.1038/s41550-021-01513-x

\bibitem[Miyachi et al.(2019)]{miyachi19} Miyachi, Y., Sakai, N., Kawata, D., et al.\ 2019, \apj, 882, 48. doi:10.3847/1538-4357/ab2f86

\bibitem[Mr{\'o}z et al.(2017)]{mroz17} Mr{\'o}z, P., Udalski, A., Skowron, J., et al.\ 2017, \nat, 548, 183. doi:10.1038/nature23276

\bibitem[Mu{\v{z}}i{\'c} et al.(2017)]{muzic17} Mu{\v{z}}i{\'c}, K., Sch{\"o}del, R., Scholz, A., et al.\ 2017, \mnras, 471, 3699. doi:10.1093/mnras/stx1906

\bibitem[Mu{\v{z}}i{\'c} et al.(2019)]{muzic19} Mu{\v{z}}i{\'c}, K., Scholz, A., Pe{\~n}a Ram{\'\i}rez, K., et al.\ 2019, \apj, 881, 79. doi:10.3847/1538-4357/ab2da4

\bibitem[Offner et al.(2014)]{offner14} Offner, S.~S.~R., Clark, P.~C., Hennebelle, P., et al.\ 2014, Protostars and Planets VI, 53. doi:10.2458/azu\_uapress\_9780816531240-ch003

\bibitem[Padoan \& Nordlund(2004)]{padoan04} Padoan, P. \& Nordlund, {\r{A}}.\ 2004, \apj, 617, 559. doi:10.1086/345413

\bibitem[Parker \& Quanz(2012)]{parker12} Parker, R.~J. \& Quanz, S.~P.\ 2012, \mnras, 419, 2448. doi:10.1111/j.1365-2966.2011.19911.x

\bibitem[Parker \& Alves de Oliveira(2017)]{parker17} Parker, R.~J. \& Alves de Oliveira, C.\ 2017, \mnras, 468, 4340. doi:10.1093/mnras/stx739

\bibitem[Parker(2020)]{parker20} Parker, R.~J.\ 2020, Royal Society Open Science, 7, 201271. doi:10.1098/rsos.201271

\bibitem[Pe{\~n}a Ram{\'\i}rez et al.(2012)]{pena12} Pe{\~n}a Ram{\'\i}rez, K., B{\'e}jar, V.~J.~S., Zapatero Osorio, M.~R., et al.\ 2012, \apj, 754, 30. doi:10.1088/0004-637X/754/1/30

\bibitem[Pontoppidan et al.(2022)]{pontoppidan22} Pontoppidan, K., Blome, C., Braun, H., et al.\ 2022, arXiv:2207.13067

\bibitem[Price \& Bate(2007)]{price07} Price, D.~J. \& Bate, M.~R.\ 2007, \mnras, 377, 77. doi:10.1111/j.1365-2966.2007.11621.x


\bibitem[Rigby et al.(2022)]{rigby22} Rigby, J., Perrin, M., McElwain, M., et al.\ 2022, arXiv:2207.05632

\bibitem[Robin et al.(2003)]{robin03} Robin, A.~C., Reyl{\'e}, C., Derri{\`e}re, S., et al.\ 2003, \aap, 409, 523. doi:10.1051/0004-6361:20031117


\bibitem[Scholz et al.(2012)]{scholz12} Scholz, A., Jayawardhana, R., Muzic, K., et al.\ 2012, \apj, 756, 24. doi:10.1088/0004-637X/756/1/24

\bibitem[Scholz et al.(2013)]{scholz13} Scholz, A., Geers, V., Clark, P., et al.\ 2013, \apj, 775, 138. doi:10.1088/0004-637X/775/2/138

\bibitem[Smith \& Bonnell(2001)]{smith01} Smith, K.~W. \& Bonnell, I.~A.\ 2001, \mnras, 322, L1. doi:10.1046/j.1365-8711.2001.04321.x

\bibitem[Spiegel \& Burrows(2012)]{spiegel12} Spiegel, D.~S. \& Burrows, A.\ 2012, \apj, 745, 174. doi:10.1088/0004-637X/745/2/174

\bibitem[Sumi et al.(2011)]{sumi11} Sumi, T., Kamiya, K., Bennett, D.~P., et al.\ 2011, \nat, 473, 349. doi:10.1038/nature10092

\bibitem[van Elteren et al.(2019)]{vanelteren19} van Elteren, A., Portegies Zwart, S., Pelupessy, I., et al.\ 2019, \aap, 624, A120. doi:10.1051/0004-6361/201834641

\bibitem[van Terwisga et al.(2020)]{vanterwisga20} van Terwisga, S.~E., van Dishoeck, E.~F., Mann, R.~K., et al.\ 2020, \aap, 640, A27. doi:10.1051/0004-6361/201937403

\bibitem[Vigan et al.(2017)]{vigan17} Vigan, A., Bonavita, M., Biller, B., et al.\ 2017, \aap, 603, A3. doi:10.1051/0004-6361/201630133

\bibitem[Vigan et al.(2021)]{vigan21} Vigan, A., Fontanive, C., Meyer, M., et al.\ 2021, \aap, 651, A72. doi:10.1051/0004-6361/202038107

\bibitem[Veras et al.(2011)]{veras11} Veras, D., Wyatt, M.~C., Mustill, A.~J., et al.\ 2011, \mnras, 417, 2104. doi:10.1111/j.1365-2966.2011.19393.x

\bibitem[Veras \& Moeckel(2012)]{verasmoeckel12} Veras, D. \& Moeckel, N.\ 2012, \mnras, 425, 680. doi:10.1111/j.1365-2966.2012.21552.x

\bibitem[Veras \& Raymond(2012)]{verasraymond12} Veras, D. \& Raymond, S.~N.\ 2012, \mnras, 421, L117. doi:10.1111/j.1745-3933.2012.01218.x

\bibitem[Wang et al.(2020)]{wang20} Wang, Y.-H., Perna, R., \& Leigh, N.~W.~C.\ 2020, \mnras, 496, 1453. doi:10.1093/mnras/staa1627

\bibitem[Widmark et al.(2022)]{widmark22} Widmark, A., Widrow, L.~M., \& Naik, A.\ 2022, arXiv:2207.03492

\bibitem[Willott et al.(2022)]{willott22} Willott, C.~J., Doyon, R., Albert, L., et al.\ 2022, \pasp, 134, 025002. doi:10.1088/1538-3873/ac5158

\bibitem[Winter et al.(2022)]{winter22} Winter, A.~J., Haworth, T.~J., Coleman, G.~A.~L., et al.\ 2022, \mnras. doi:10.1093/mnras/stac1564

\bibitem[Wittenmyer et al.(2020)]{wittenmyer20} Wittenmyer, R.~A., Wang, S., Horner, J., et al.\ 2020, \mnras, 492, 377. doi:10.1093/mnras/stz3436

\bibitem[Wolthoff et al.(2022)]{wolthoff22} Wolthoff, V., Reffert, S., Quirrenbach, A., et al.\ 2022, \aap, 661, A63. doi:10.1051/0004-6361/202142501

\bibitem[Wright et al.(2012)]{wright12} Wright, J.~T., Marcy, G.~W., Howard, A.~W., et al.\ 2012, \apj, 753, 160. doi:10.1088/0004-637X/753/2/160

\bibitem[Wurster et al.(2019)]{wurster19} Wurster, J., Bate, M.~R., \& Price, D.~J.\ 2019, \mnras, 489, 1719. doi:10.1093/mnras/stz2215


\bibitem[Zheng et al.(2015)]{zheng15} Zheng, X., Kouwenhoven, M.~B.~N., \& Wang, L.\ 2015, \mnras, 453, 2759. doi:10.1093/mnras/stv1832

\end{thebibliography}
\end{document}